\input harvmac

\def \ep{\epsilon}
\def\D{\Delta}

\def \del {\partial}

\def \ha{{\textstyle{1\over 2}}}

\def \D {\Delta}
\def \a {\alpha}

\def \chi {\chi}
\def \s {\sigma}
\def \p {\phi}
\def \m {\mu}
\def \n {\nu}

\def \td {\tilde }

\def \sm {$\s$-model }

\def \inv {^{-1}}
\def \ov {\over }

\def \D {\Delta}

\def \lr { \lref}
\def\np {{  Nucl. Phys. }}
\def \pl {{  Phys. Lett. }}
\def \mpl {{ Mod. Phys. Lett. }}
\def \prl {{  Phys. Rev. Lett. }}
\def \pr  {{ Phys. Rev. }}

\def \cqg {{ Class. Quant. Grav. }}

\baselineskip8pt
\Title{
\vbox
{\baselineskip 6pt{\hbox{  }}{\hbox
{Imperial/TP/95-96/38 }}{\hbox{hep-th/9604035}} {\hbox{
  }}} }
{\vbox{\centerline {Harmonic superpositions of M-branes}
}}
\vskip -20 true pt
\medskip
\medskip
\centerline{   A.A. Tseytlin\footnote{$^{\star}$}{\baselineskip8pt
e-mail address: tseytlin@ic.ac.uk}\footnote{$^{\dagger}$}{\baselineskip8pt
On leave  from Lebedev  Physics
Institute, Moscow.} }

\smallskip\smallskip
\centerline {\it  Theoretical Physics Group, Blackett Laboratory,}
\smallskip

\centerline {\it  Imperial College,  London SW7 2BZ, U.K. }
\bigskip\bigskip
\centerline {\bf Abstract}
\medskip
\baselineskip10pt
\noindent
We present solutions describing supersymmetric configurations of 2 or 3 orthogonally intersecting 2-branes and 5-branes of $D=11$ supergravity.
The configurations which preserve  1/4 or 1/8  of maximal supersymmetry are
 $2\bot 2,\ 5\bot 5,\ 2\bot 5,$ $\ 2\bot 2 \bot 2,\ 5 \bot 5 \bot 5,\  2 \bot 2 \bot 5$ and  $ \ 2  \bot 5 \bot 5$
($2\bot 2$ stands for orthogonal intersection of  two 2-branes over a point, etc.;  p-branes of the same type intersect over (p-2)-branes).
There exists a simple rule which governs the construction of 
composite supersymmetric p-brane solutions in $D=10$ and $11$ 
with a separate  harmonic function assigned to each constituent  
1/2-supersymmetric p-brane. The resulting picture of intersecting p-brane 
solutions complements their D-brane interpretation in $D=10$ and seems to
support possible existence of a $D=11$ analogue of D-brane description.
The $D=11$ solution describing intersecting 2-brane and 5-brane 
reduces in $D=10$ to a type II string solution corresponding to a 
fundamental string lying within a solitonic 5-brane (which further 
reduces to an extremal $D=5$ black hole). 
We also discuss a particular  $D=11$ embedding of the extremal 
$D=4$ dyonic black hole solution with finite area of horizon.

\medskip
\Date {April 1996}
\noblackbox
\baselineskip 14pt plus 2pt minus 2pt
\lr \dgh {A. Dabholkar, G.W. Gibbons, J. Harvey and F. Ruiz Ruiz,  \np
B340 (1990) 33;
A. Dabholkar and  J. Harvey,  \prl
63 (1989) 478.
}
\lr\mon{J.P. Gauntlett, J.A. Harvey and J.T. Liu, \np B409 (1993) 363.}
\lr\chs{C.G. Callan, J.A. Harvey and A. Strominger, 
\np { B359 } (1991)  611.}

\lr \CM{ C.G. Callan and  J.M.  Maldacena, 
PUPT-1591,  hep-th/9602043.} 
\lr\SV {A. Strominger and C. Vafa, HUTP-96-A002,  hep-th/9601029.}

\lr\MV {J.C. Breckenridge, R.C. Myers, A.W. Peet  and C. Vafa, HUTP-96-A005,  hep-th/9602065.}

\lr \CT{M. Cveti\v c and  A.A.  Tseytlin, 
\pl { B366} (1996) 95, hep-th/9510097. 
}
\lr \CTT{M. Cveti\v c and  A.A.  Tseytlin, 
IASSNS-HEP-95-102, hep-th/9512031. 
}
\lr\LW{ F. Larsen  and F. Wilczek, 
PUPT-1576,  hep-th/9511064.    }
\lr\TT{A.A. Tseytlin, \mpl A11 (1996) 689,   hep-th/9601177.}
\lr \HT{ G.T. Horowitz and A.A. Tseytlin,  \pr { D51} (1995) 
2896, hep-th/9409021.}
\lr\khu{R. Khuri, \np B387 (1992) 315; \pl B294 (1992) 325.}
\lr\CY{M. Cveti\v c and D. Youm,
 UPR-0672-T, hep-th/9507090; UPR-0675-T, hep-th/9508058; 
  \pl { B359} (1995) 87, 
hep-th/9507160.}

\lr\ght{G.W. Gibbons, G.T. Horowitz and P.K. Townsend, \cqg 12 (1995) 297,
hep-th/9410073.}
\lr\dul{M.J. Duff and J.X. Lu, \np B416 (1994) 301, hep-th/9306052. }
\lr\hst {G.T. Horowitz and A. Strominger, hep-th/9602051.}
\lr\dull{M.J. Duff and J.X. Lu, \pl B273 (1991) 409. }
\lr \guv{R. G\"uven, \pl B276 (1992) 49. }
\lr \gups {S.S. Gupser, I.R.   Klebanov  and A.W. Peet, 
hep-th/9602135.}
\lr \dus { M.J. Duff and  K.S. Stelle, \pl B253 (1991) 113.}

\lr\hos{G.T.~Horowitz and A.~Strominger, Nucl. Phys. { B360}
(1991) 197.}
\lr\teit{R. Nepomechi, \pr D31 (1985) 1921; C. Teitelboim, \pl B167 (1986) 69.}
\lr \duf { M.J. Duff, P.S. Howe, T. Inami and K.S. Stelle, 
\pl B191 (1987) 70. }
\lr\duh {A. Dabholkar and J.A. Harvey, \prl { 63} (1989) 478;
 A. Dabholkar, G.W.   Gibbons, J.A.   Harvey  and F. Ruiz-Ruiz,
\np { B340} (1990) 33. }
\lr\mina{M.J. Duff, J.T. Liu and R. Minasian, 
\np B452 (1995) 261, hep-th/9506126.}
\lr\dvv{R. Dijkgraaf, E. Verlinde and H. Verlinde, hep-th/9603126.}
\lr\gibb{G.W. Gibbons and P.K. Townsend, \prl  71
(1993) 3754, hep-th/9307049.}
\lr\town{P.K. Townsend, hep-th/9512062.}
\lr\kap{D. Kaplan and J. Michelson, hep-th/9510053.}
\lr\hult{
C.M. Hull and P.K. Townsend, Nucl. Phys. { B438} (1995) 109;
P.K. Townsend, Phys. Lett. {B350} (1995) 184;
E. Witten, \np B443 (1995) 85; 
J.H. Schwarz,  \pl B367 (1996) 97, hep-th/9510086, hep-th/9601077;
P.K. Townsend, hep-th/9507048;
M.J. Duff, J.T. Liu and R. Minasian, 
\np B452 (1995) 261, hep-th/9506126; 
K. Becker, M. Becker and A. Strominger, Nucl. Phys. { B456} (1995) 130;
I. Bars and S. Yankielowicz, hep-th/9511098;
P. Ho{\v r}ava and E. Witten, Nucl. Phys. { B460} (1996) 506;
E. Witten, hep-th/9512219.}
\lr\beck{
K. Becker and  M. Becker, hep-th/9602071.}
\lr\aar{
O. Aharony, J. Sonnenschein and S. Yankielowicz, hep-th/9603009.}
\lr\ald{F. Aldabe, hep-th/9603183.}
\lr\ast{A. Strominger, hep-th/9512059.}
\lr \ttt{P.K. Townsend, hep-th/9512062.}
\lr \papd{G. Papadopoulos and P.K. Townsend, hep-th/9603087.}
\lr\jch {J. Polchinski, S. Chaudhuri and C.V. Johnson, 
hep-th/9602052.}
\lr \ddd{E. Witten, hep-th/9510135;
M. Bershadsky, C. Vafa and V. Sadov, hep-th/9510225;
A. Sen, hep-th/9510229, hep-th/9511026;
C. Vafa, hep-th/9511088;
M. Douglas, hep-th/9512077. }

\lr \gig{G.W. Gibbons, M.J. Green and M.J. Perry, 
hep-th/9511080.}

\lr \dufe{M.J. Duff, S.  Ferrara, R.R. Khuri and 
J. Rahmfeld, \pl B356 (1995) 479,  hep-th/9506057.}

\lr\stp{H. L\" u, C.N. Pope, E. Sezgin and K.S. Stelle, \np B276 (1995)  669, hep-th/9508042.}
\lr \duff { M.J. Duff and J.X. Lu, \np B354 (1991) 141. } 
\lr \pol { J. Polchinski, \prl 75 (1995) 4724,  hep-th/9510017.} 
\lr \iz { J.M. Izquierdo, N.D. Lambert, G. Papadopoulos and 
P.K. Townsend,  \np B460 (1996) 560, hep-th/9508177. }

\lr \US{M. Cveti\v c and  A.A.  Tseytlin, 
\pl {B366} (1996) 95, hep-th/9510097;   hep-th/9512031.  
}
\lr\mast{J. Maldacena and A. Strominger, hep-th/9603060.}
\lr \CY{M. Cveti\v c and D. Youm,
 \pr D53 (1996) 584, hep-th/9507090.  }
 \lr\kall{R. Kallosh, A. Linde, T. Ort\' in, A. Peet and A. van Proeyen, \pr { D}46 (1992) 5278.} 
\lr \grop{R. Sorkin, Phys. Rev. Lett. { 51 } (1983) 87;
D. Gross and M. Perry, Nucl. Phys. { B226} (1983) 29. 
}
\lr \myers{C. Johnson, R. Khuri and R. Myers, hep-th/9603061.} 

\lr\TTT{I.R. Klebanov and A.A. Tseytlin, ``Intersecting M-branes as four dimensional black holes",  hep-th/9604166. }
\lr \green{M.B. Green and M. Gutperle, hep-th/9604091.}
\lr \myk{R.R. Khuri and R.C. Myers, hep-th/9512061.}
\lr\lup {H. L\" u and C.N. Pope, hep-th/9512012; hep-th/9512153.}

\lr \wal {D. Garfinkle, \pr D46 (1992) 4286.
 }

\lr \berg{E. Bergshoeff, C. Hull and T. Ortin, \np B451 (1995) 547, hep-th/9504081.}

\newsec{Introduction}
In view of  recent suggestions that  $D=11$ supergravity 
may be a  low-energy effective field theory of  
a fundamental `M-theory'  which generalises  known string theories 
(see,  e.g.,  \refs{\hult})
it is  important to gain better understanding 
of its   classical p-brane solutions.
It seems likely that  supersymmetric BPS saturated 
p-brane solutions of low-dimensional 
theories  can be understood as  `reductions' 
 of basic $D=11$  `M-branes' -- 2-brane \dus\ 
and 5-brane \guv\ and their combinations \papd. 
The important questions are    which combinations
of M-branes  
do  actually  appear as  stable supersymmetric 
solutions, how  to construct them and how they are related to 
similar $D=10$ p-brane  configurations.
 
Here we  shall follow and extend further the suggestion  \papd\ 
 that  stable supersymmetric $D=11$ p-brane    configurations
should  have an interpretation  in terms of orthogonal intersections 
of  certain numbers of 2-branes and/or  5-branes.
A possibility of existence of similar supersymmetric configurations 
 was   pointed out earlier (on the basis of 
charge conservation and supersymmetry  considerations) in \refs{\ast,\ttt}.
Discussions of related systems of  D-branes in $D=10$ 
string theories appeared in 
\refs{\ddd, \jch,\green}.

It should be noted that  `intersecting p-brane' solutions in \papd\
and below are isometric in all directions internal to all constituent p-branes
(the background fields depend only on the remaining common
transverse directions). They are different from possible
 virtual configurations 
where, e.g., a (p-2)-brane ends  (in transverse space radial direction)
on a p-brane \ast\ (such  configurations may
contribute to path integral but may  not correspond to stable classical solutions). A configuration of, e.g.,
  a p-brane  and a p$'$-brane  intersecting in  p+p$'$-space
 may be also considered as 
a  special anisotropic (cf.\myk)   p+p$'$-brane.
 We expect (see also \papd)  that there should exist
more general solutions (with constituent p-branes effectively
having different transverse
spaces)  which represent more complicated `BPS  bound states' 
of constituent p-branes and interpolate 
between  such intersecting solutions
and solutions  with higher rotational symmetry for 
each p-brane.

The basic property of  supersymmetric p-brane solutions 
of supergravity theories is that they are expressed  in terms of 
harmonic functions  of transverse spatial coordinates. This 
reflects the BPS saturated nature of these solutions and implies that 
there exist stable `multicenter'  configurations of multiple  parallel p-branes
of the same type.  There may  also exist stable 
supersymmetric solutions corresponding to
combinations (intersections and bound states) 
of p-branes of the  same or different types.
While the rules of combining p-branes 
(in a way preserving supersymmetry and charge conservation) 
in $D=10$ depend on a type (NS-NS or R-R) of the constituents  
 \ast, 
the following rules seem to  be universal in  $D=11$
(these rules are consistent with $D=10$ rules 
upon dimensional reduction):\foot{Related conditions  for supersymmetric combinations of D-branes 
in $D=10$ are  that the number of mixed Dirichlet-Neumann directions 
should be a multiple of 4 and that a (p-2)-brane can lie 
within a p-brane  \jch.}

\noindent
(i) p-branes of the same type can intersect only over a (p-2)-brane  \papd\ (i.e.  2-branes can intersect  over a  0-brane,  5-branes
can intersect  over a 3-brane,   3-branes 
 can intersect over a string);

\noindent
(ii) 2-brane can  orthogonally intersect 5-brane over a string \refs{\ast,\ttt};

\noindent
(iii) a configuration of $n$  
orthogonally intersecting  M-branes  preserves at least 
$1/2^{n}$ of maximal supersymmetry.\foot{In the case of  general 
solutions involving  parallel families  of p-branes 
$n$  stands for a number of   intersecting  families.}
    
Thus  in addition to the  basic (2-  and  5-)  M-branes    
 preserving 1/2 of supersymmetry  
one should  expect to find also the following 
composite configurations:

\noindent
 (i)  $2\bot 2$, \  $5\bot 5$, \
$5\bot 2$     
  preserving 1/4 of supersymmetry, and

\noindent 
 (ii) $2\bot 2\bot 2$, \ $5\bot 2\bot 2$,\ $5\bot 5\bot 2$,\ 
$5\bot 5\bot 5$ 
 preserving 1/8 of supersymmetry.

The allowed  1/16 supersymmetric configurations with 
{\it four}  intersecting M-branes 
  (i.e. $2\bot 2\bot 2\bot 2 $, \  $ 2\bot 2\bot 2\bot 5$,\ 
$5\bot 5\bot 5\bot 2 $)
 have transverse space dimension $d  < 3$ 
and thus  (being described in terms 
of  harmonic functions  of transverse coordinates) 
 are not asymptotically flat in transverse directions.
The exception is  $5\bot 5\bot 2\bot 2$
for which the transverse dimension is 3 
as in the  $5\bot 2\bot 2$,\ $5\bot 5\bot 2$ and 
$5\bot 5\bot 5$  cases.
Like the `boosted' version of  $5\bot 5\bot 5$  solution 
the $5\bot 5\bot 2\bot 2$
background is 1/8-supersymmetric 
and upon compactification  to $D=4$ reduces 
to the dyonic $D=4$ black hole \refs{\CY,\US}  with four different charges  
and finite area of the horizon. 
This will be discussed in  detail in \TTT. 
Note also that the regular 3-charge dyonic $D=5$ black hole \TT\
is described  by  $2\bot 2\bot 2$ or by `boosted'   $2\bot 5$  solution.

In \papd\ the  `electric' $D=11$ solutions of \guv\ with 
1/4 and 1/8 of supersymmetry were interpreted  as  special 
$2\bot 2$ and $2\bot 2\bot 2$ configurations
and the corresponding  `magnetic' $5\bot 5$ and $5\bot 5\bot 5$
solutions  were found. 

Below we shall generalise the  solutions  of \refs{\guv,\papd}
to the case when each  intersecting p-brane is described by a 
separate  harmonic function 
and will also  present new solutions corresponding to case when intersecting 
M-branes are of different type, i.e.  $5\bot 2$, $5\bot 2\bot 2$ and  $5\bot 5\bot 2$.  The important  $5\bot 2$
solution reduces in $D=10$ to a   configuration 
which can be interpreted as a   fundamental string lying  within  a
  solitonic  (i.e. NS-NS) 5-brane  (such $D=10$ solution was given in \TT).\foot{In addition to the intersecting $2\bot 5$ configuration
there should  exist a supersymmetric  $D=11$ 
solution describing a 2-brane lying within a  5-brane (see \iz\ and Section 3.2). 
 It should  lead upon dimensional 
reduction  (along 5-brane direction orthogonal to 2-brane)
to a 2-brane within 4-brane   configuration
of type IIA theory (related   by $T$-duality to a  R-R string within 
3-brane in type IIB theory) 
which  is allowed  from the point of view of D-brane description \jch. } 

The basic  observation that clarifies  
the picture  suggested in \papd\ 
and  leads to  various generalisations (both in $D=11$ and $D=10$)
is that it is possible to assign an independent  
harmonic function to each intersecting p-brane (the solutions 
in \refs{\guv,\papd} correspond to the  `degenerate'  case when all harmonic 
functions are taken to be equal).
For example, a generalisation of  $2\bot 2$ solution of \refs{\guv,\papd}
now parametrised  by two independent harmonic functions
describes, in particular,  two orthogonally
 intersecting families of parallel 2-branes. 

Combining the above $D=11$ p-brane composition  rules with the 
 `harmonic function rule' explained 
and illustrated on $D=10$ examples
in Section 2 below,  
 it is easy to write down   explicitly   new  solutions  
representing  orthogonally intersecting (parallel families of) 
2-branes  and 5-branes mentioned above, i.e.
$5\bot 2$,  $5\bot 2\bot 2$,  $5\bot 5\bot 2$ (Section 3).
A special  version of $2\bot 5$  solution
superposed with a Kaluza-Klein monopole 
 represents  a particular  $D=11$ embedding 
of the extreme  dyonic $D=4$  black hole  (Section 4).

\newsec{Harmonic function rule and  $D=10$ intersecting  p-brane solutions}
The metric and 4-form field strength of the 
basic  extremal supersymmetric $p=2$  \dus\ and $p=5$  \guv\ 
p-brane solutions of $D=11$ supergravity can be represented in the following form 
\eqn\two{
d s^2_{11} =  
H^{(p+1)/9}_p (x) \big[ H\inv_p (x)(-dt^2 + dydy_{p}) +  dxdx_{10-p}\big] \ ,    }
\eqn\ff{
{{\cal F}_4}_{(2)} = -3dt\wedge d(H_2\inv J) \ , \ \ \ \ \ 
{{\cal F}_4}_{(5)} = 3 *dH_5 \ ,  \ \ \ \  \  \del^2 H_p=0 \ , }
where $dydy_{p} \equiv  dy_1^2 + ... +dy_p^2, \
$  $dxdx_{n} \equiv  dx_1^2 + ... +dx_{n}^2$ ($y_a$ are internal coordinates of $p$-brane and $x_i$ are transverse coordinates), 
$J=dy_1 \wedge dy_2$ is the volume  form on $R^2_y$  
 and $*$ defines the  dual form  in $R^5_x$.
$H_p$ is a harmonic function on  $R^{10-p}_x$ which 
 may  depend only on part of $x$-coordinates
(this may  be viewed as a result of taking a periodic array 
of generic 1-center solutions;  
for simplicity, we shall still refer to  such  solution as a  p-brane
even though it will be `delocalised' in some $x$-directions). 

The structure of ${{\cal F}_4}$ in \ff\  is such that 
the contribution of the CS interaction term to the ${{\cal F}_4}$ - equation
of motion vanishes (i.e. ${{\cal F}_4} \wedge {{\cal F}_4}=0$).
This will also  be the property  of all intersecting solutions discussed below.

The structure  of the metric \two\ can be  described as follows. 
If one separates the overall conformal factor  which multiplies
the transverse  $x$-part then each of the squares of differentials 
of the coordinates belonging to a given 
 p-brane is multiplied by the inverse power 
of the corresponding harmonic function. 
We suggest  that this as  a  general rule (`{\it harmonic function rule}')
which applies  to any supersymmetric
combination of orthogonally intersecting  p-branes: 
if the   coordinate 
$y$ belongs to several constituent  p-branes ($p_1,...,p_n$) 
then its contribution to  the metric written in the conformal frame where 
the transverse part  $dxdx$ is `free'  is 
multiplied by the product of the inverse powers of harmonic functions corresponding to each of the  p-branes it belongs to, i.e. 
$H\inv_{p_1} ... H\inv_{p_n} dy^2$.
The harmonic function factors thus play the role of  `labels' of 
 constituent p-branes 
 making  the interpretation of the metric 
straightforward.

It can be  checked explicitly that the specific 
backgrounds discussed below 
which  can be constructed using this rule 
 indeed solve the $D=11$ supergravity equations of motion.
While we did  not attempt to give a  general derivation of this rule
directly from $D=11$  field equations,   it should be 
a consequence of the fact that intersecting configurations 
are required to be supersymmetric (i.e. it should follow from first-order 
equations implied by the existence of a Killing spinor).
 Since one should be 
able to superpose BPS states 
they must be parametrised (like their basic constituent p-branes) by 
harmonic functions. 
Taking the centers of 
each of the harmonic function at different points one can 
interpolate between the cases of far separated and coinciding  p-branes, confirming
the consistency of the `harmonic function rule'.

This rule is also consistent (upon dimensional reduction) 
with analogous one which operates in $D=10$  where it can be justified 
 by conformal \sm considerations (for specific  NS-NS configurations) \refs{\US,\TT}
or by $T$-duality \berg\  considerations (for R-R configurations).

\subsec{ $2\bot 2$ $D=11$ solution }
For example, the metric of two  $D=11$ 2-branes intersecting 
over a point   constructed according to the above  rule will be 
\eqn\twot{
d s^2_{11} =  
H^{1/3}_{2(1)} (x) H^{1/3}_{2(2)} (x) 
\big[ - H\inv_{2(1)} (x) H\inv _{2(2)} (x) dt^2  } $$
+ \  H\inv_{2(1)}(x) dydy_{2}^{(1)}  + H\inv _{2(2)}(x) 
 dydy_{2}^{(2)}  +  dxdx_{6}\big] \ .    $$
Here $y^{(1)}_1,y^{(1)}_2$ and $y_1^{(2)},y_2^{(2)}$
 are  internal coordinates 
of the two 2-branes.  
The time part `belongs' to  both  2-branes 
and thus is multiplied by the product of the inverse 
powers of both harmonic functions.
  The corresponding field strength is 
\eqn\fff{ {{\cal F}_4}_{(2\bot 2)} = -3dt\wedge d(H_{2(1)} \inv J_1 + H\inv_{2(2)} J_2) \ . }
The fact that the two harmonic functions 
can be centered at different (e.g. far separated) 
points  together with  supersymmetry  and  exchange 
symmetry  with respect to the two 2-branes
uniquely determines the form of the background, 
which indeed solves the $D=11$ supergravity equations.

Setting $H_{2(2)}=1$ 
 one  gets back  to the special 2-brane 
solution \two,\ff\ where $H_2=H_{2(1)}$ does not depend on two of the eight $x$-coordinates
(called $y^{(2)}$ in \twot).
Another special case  $H_{2(1)}=H_{2(2)}$ 
corresponds to the  `4-brane' solution of \guv\ interpreted in \papd\ as representing 
two 
intersecting 2-branes. 
 
\subsec{Examples of intersecting  p-brane 
   solutions  of  $D=10$ type II  theories }
Before proceeding with the discussion of  other composite 
$D=11$ solutions let us demonstrate  how  the `harmonic function rule'
applies  to various  p-brane solutions 
of $D=10$  type II superstring theories. 

The basic $D=10$ fundamental string solution  \duh\
which has the following metric (we  shall always 
use the string-frame  form of the $D=10$ metric)
\eqn\fus{
ds^2_{10} = H^{-1}_1(x)  (-dt^2 + dy^2) + dxdx_{8} \ . }
The metric of the solution 
describing  a  fundamental string lying  within   
the solitonic  5-brane  \refs{\chs,\duff}  is given by  \TT\
\eqn\fuss{
ds^2_{10} = H^{-1}_1 (x) (-dt^2 + dy^2_1) +  dy^2_2 + ....+dy^2_5 + 
     H_5 (x)dxdx_{4} }  $$  =  \ 
H_5 (x) \big[ H^{-1}_1 (x) H\inv _5 (x) (-dt^2 + dy^2_1)
         +  H\inv _5 (x) (dy^2_2 + ....+dy^2_5) + dxdx_{4} \big] \ .  $$
Other NS-NS background fields have obvious `direct sum' structure, i.e.
the  dilaton is given by $e^{2\p} = H_1\inv H_5$
and the antisymmetric  2-tensor   has both `electric' (fundamental string)  and `magnetic' (5-brane) components,  
$B_{ty_1}= H^{-1}_1 , \ H_{mnk}= -\ep_{mnkl}\del_l H_5$.
The factorised harmonic function structure of this background has a natural 
explanation from the point of view of the  associated  conformal 
$\s$-model \TT. 
The  solutions \fus,\fuss\  (as well as all  solutions
below which have a null hypersurface-orthogonal isometry)
admit a straightforward `momentum along  string' generalisation 
$-dt^2 + dy^2_1 \to  -dt^2 + dy^2_1 + K(x) (dt-dy_1)^2$
where $K$ is an independent harmonic function (cf. \wal).

Applying $SL(2,Z)$ duality transformation  of type IIB supergravity 
(which  inverts the dilaton and 
does not change the Einstein-frame metric, i.e.  
modifies   the string  frame metric only by the conformal factor
$e^{-\p}$)
one learns  that the metric describing  a R-R  string lying  within  a R-R 5-brane
has the same structure as \fuss,  i.e. the structure consistent with the 
harmonic function rule (with the  factor multiplying the square bracket now being 
$ H^{1/2}_1  H^{1/2} _5$). 
$T$-duality in  the two 5-brane directions orthogonal 
to the string gives  type IIB solution describing two 3-branes
orthogonally intersecting  over a string. Its metric  
has the form consistent 
with the `harmonic function rule' 
 \eqn\fwww{
ds^2_{10}=
H^{1/2}_{3(1)}  H^{1/2}_{3(2)}  
\big[ H\inv_{3(1)} H\inv _{3(2)} (- dt^2  + dy_1^2) } $$
+ \  H\inv_{3(1)} dydy_{2}^{(1)}
  + H\inv _{3(2)} dydy_{2}^{(2)}  +  dxdx_{4}\big] \ ,     $$
where $y_1$ is the coordinate common to the two 3-branes.\foot{Adding a boost along the common string one finds upon reduction to $D=5$ an 
extremal black hole with 3 charges and  $x=0$ as a regular horizon \TT.} 
The corresponding self-dual 5-tensor is 
\eqn\fer{
{{\cal F}_5}_{(3\bot 3)}= dt\wedge (dH\inv_{3(1)} \wedge dy_1 \wedge dy_2^{(1)} \wedge dy_3^{(1)}
+  dH\inv_{3(2)} \wedge dy_1 
\wedge dy_2^{(2)} \wedge dy_3^{(2)}) } $$ 
 +\   *d H_{3(1)}\wedge dy_2^{(2)} \wedge dy_3^{(2)})
+ *dH_{3(2)}\inv \wedge dy_2^{(1)} \wedge dy_3^{(1)} \ . $$
More general 1/8 supersymmetric solutions describing the 
configurations 
$3\bot 3\bot 3$ and $3\bot 3\bot 3\bot 3$
 will be discussed in \TTT.

While charge conservation  prohibits the   configuration 
with a fundamental string orthogonally intersecting solitonic 5-brane
(and, by $SL(2,Z)$ duality,  R-R string intersecting R-R 5-brane), 
 the type IIB  configuration  of   a fundamental   string 
intersecting a  R-R  5-brane 
(and its dual --   R-R string intersecting 
 a solitonic  5-brane) 
is   allowed  \ast.
 The corresponding solution is   straightforward to write down. Its    metric  is  given by (cf. \fuss; see also the discussion below)
\eqn\fusst{
ds^2_{10} =
H_5^{1/2} (x) \big[ - H^{-1}_1 (x) H\inv _5 (x) dt^2 +  H_1\inv dy^2_1 } 
  $$        + \  H\inv _5 (x) (dy^2_2 + ....+dy^2_6) + dxdx_{3} \big] \ .    $$ 
Here $y_1$ is the coordinate of the string intersecting 5-brane ($y_2,...,y_6$) over 
a point.\foot{For a multicenter choice of 5-brane harmonic function $H_5$ 
this metric describes  a
fundamental string intersecting several parallel 
5-branes.}

In general,  metrics of  1/2-supersymmetric 
 p-branes of type II theories which carry   R-R  charges  
have the following form  \refs{\hos}
\eqn\fef{ ds^2_{10} 
=   H^{1/2}_p[ H\inv_p (-dt^2 + dydy_p)  +   dxdx_{9-p}]  \ , }
with the  dilaton  given by  $e^{2\p} = H^{(3-p)/2}_p$.
It is  straightforward to apply the  `harmonic function rule'
and  the supersymmetry and R-R charge conservation  rules  \ast\ 
to construct explicitly the  solutions  which  describe 
multiple and intersecting R-R soliton configurations 
which are counterparts of the D-brane configurations  discussed in
 \refs{\ddd,\jch}.
The  resulting  procedure of  constructing `composite'  supersymmetric 
backgrounds from `basic'  ones is   in  direct
 correspondence with a  picture   of `free' parallel or intersecting 
D-brane hypersurfaces  in flat space \pol.

For example, 
the solution  of type IIB theory representing  a   R-R string  orthogonally intersecting 
3-brane ($T$-dual to a  0-brane within a 4-brane 
 in type IIA theory) is described by 
\eqn\fof{ ds^2_{10} 
=   H^{1/2}_1 H^{1/2}_3 [- H\inv_1 H\inv_3 dt^2 +  H\inv_1  dy_1^2  
 +  H\inv_3 (dy_2^2 + dy_3^2 + dy_4^2) +   dxdx_{5}]  \ .  } 
By  $SL(2,Z)$ duality  the  same  (up to a conformal factor) metric
represents   a fundamental 
string intersecting  a 3-brane. 

An example of intersecting solution in type IIA theory is 
provided by  a fundamental string orthogonally intersecting a  4-brane 
at a point (cf. \fuss,\fof) 
\eqn\jjj{ ds^2_{10} 
=    H^{1/2}_4 \big[- H\inv_1 H\inv_4 dt^2 +  H\inv_1  dy_1^2  
 +  H\inv_4 (dy_2^2 + dy_3^2 + dy_4^2 + dy_5^2) +   dxdx_{4}\big]  \ .  }
The required  dilaton and   antisymmetric tensors 
are  given by  direct sums  of constituent fields. 
This background will be reproduced  in Section 3.2 by  dimensional reduction
of  orthogonally intersecting 2-brane and 5-brane  solution 
of  $D=11$ supergravity.  

Metrics describing  configurations  of 
different  parallel 
type II p-branes  lying within each other 
(with at least one of them being  of R-R type)  
do not  obey  the `harmonic function rule'. For example, 
the metric of the `fundamental string -- R-R string' bound state solution
of type IIB theory  (obtained by applying 
$SL(2,Z)$ transformation to the fundamental string background 
 \fef, see Schwarz in \hult)
has the following structure 
\eqn\fouf{ ds^2_{10} 
=   \td H^{1/2}_1 [ H\inv_1 (-dt^2 +  dy_1^2 )  
      +   dxdx_{8}]  \ ,   } 
where  $ H_1$  and $\td H_1$ are 1-center harmonic functions
with charges $q$ and $\td q =q d^2 /(c^2 + d^2)$. 
The  
fundamental string limit  corresponds to $\td H_1=1$ while the pure 
R-R string 
is recovered when  $\td H_1=H_1$. 
Other solutions related by $T$ and $SL(2,Z)$ dualities  (e.g.
R-R string lying within 3-brane)  have similar structure.

Let us note  also that there exist a  more general class of  p-brane solutions
\refs{\dul,\ght,\stp} of the equations following from  the 
action  $S=  \int d^D x \sqrt g [ R - \ha (\del \p)^2 -  {1\ov 2 (D-2-p)!} e^{-a\p} F^2_{D-2-p}]$,  with the metric being 
\eqn\fefr{ ds^2_{D} 
=   H^{\a}_p\big[ H^{-N}_p (-dt^2 + dydy_p)  +   dxdx_{D-1-p}\big ]  \ , }
$$ N= {4\over \D} \  ,  \ \ \   \a = { 4(p+1) \ov (D-2 )  \D}  , \  \ \ \ 
\D\equiv  a^2 +  2(p+1) {D-3-p  \ov D-2 } \ . $$
The power  $N$ is integer for  supersymmetric   
p-branes   with the amount of residual supersymmetry being at least  $1/2^N$ 
of maximal (for $N=4$  and $D=4+p$ the remaining fraction 
 of supersymmetry is 1/8).
Lower  dimensional ($D < 10$)  solutions 
which have  $N >1$ 
can be re-interpreted as special limits of (reductions of) combinations  of 
1/2-supersymmetric `basic'  ($N=1$) p-brane solutions in $D=10,11$.
The higher than first  power of the $H\inv_p$ factor in the square bracket
in \fefr\ is  a result of identifying  the 
harmonic functions  corresponding to   basic 
constituent  p-branes \lup.

An example of a solution with $N=2$ is the self-dual string   
in $D=6$ \dul. It   indeed can be  reproduced   as a special limit 
of the solitonic 5-brane plus fundamental string solution \fuss\ with 
  the   four `extra' 5-brane directions wrapped around 
a 4-torus  (leading to  the solution equivalent to the 
 dyonic string of \dufe)  and  the harmonic functions  
$H_1$ and $H_5$ set equal to each other.

\newsec{Intersecting 2-branes and 5-branes in $D=11$}

\subsec{$2\bot 2\bot 2$ and $5\bot 5 \bot 5$ configurations}
To write down the explicit form of intersecting 
2- and 5-brane   solutions  in  $D=11$ 
it is useful first to simplify the notation: we shall use 
$T$ ($F$) to denote the inverse power of harmonic function corresponding to a two-brane (five-brane), 
i.e. $T\equiv H_2\inv, \ F \equiv H_5\inv$. The lower
index on $T$ or $F$ will 
indicate a number of a p-brane.
 
The solution which describes  three 2-branes intersecting over a point 
is given by the straightforward generalisation of \twot,\fff\
\eqn\twoty{
d s^2_{11} =  (T_1 T_2 T_3)^{-1/3}
 \big[ - T_1 T_2 T_3  dt^2  } $$
+ \  T_1 dydy^{(1)}_{2}  +   T_2 dydy^{(2)}_{2} +  T_3 dydy^{(3)}_{2}  
+  dxdx_{4}\big] \ ,     $$
\eqn\ffrf{ {{\cal F}_4}_{(2\bot 2\bot 2)}
 = -3dt\wedge d(T_1 J_1 + T_2 J_2  + T_3 J_3) \ .  }
The three 2-branes are parametrised by  3 sets of coordinates $y^{(i)}_{1}, y^{(i)}_{2}$
and $J_i$ are the volume forms on the corresponding 2-planes. Also, 
$\del^2 T_i\inv=0$, i.e.  $T_i\inv  = 1 + q_i/|x|^2$
in the simplest 1-center case.
The special case of $T_1=T_2=T_3$ 
gives  the `6-brane' solution of \guv\ 
correctly interpreted in \papd\ 
as representing   three
  2-branes orthogonally intersecting at one point.  
Other obvious  special choices, e.g. $T_3=1$,  
lead to a  particular case of $2\bot 2$ solution \twot,\fff\
with the harmonic functions do not depending  on two 
of the transverse coordinates. 

This solution is regular at $x=0$ and 
upon dimensional reduction to $D=5$ along $y_n$-directions
it becomes the 3-charge $D=5$ Reissner-Nordstr\"om type black hole (discussed in the special case 
of equal charges in \SV) 
which is U-dual to NS-NS dyonic black hole constructed in \TT.

Similar generalisation of the  $5\bot 5\bot 5$ 
solution in \papd\  corresponding to the three 5-branes  intersecting 
pairwise over 3-branes which in turn intersect over a string
 can be   found  by applying the `harmonic function rule'
\eqn\fivv{
d s^2_{11} =  (F_1 F_2 F_3)^{-2/3}
 \big[ F_1 F_2 F_3 (-  dt^2  + dy_0^2)  } $$
+ \  F_2 F_3 dydy^{(1)}_{2}  + 
  F_1 F_3  dydy^{(2)}_{2} +  F_1 F_2 dydy^{(3)}_{2}  
+  dxdx_{3}\big] \ ,     $$
\eqn\ffrof{ {{\cal F}_4}_{(5\bot 5\bot 5)}
 =3( *dF\inv_1 \wedge J_1 +  *dF\inv_2 \wedge J_2 +  *dF\inv_3 \wedge J_3) \ .  }
The coordinate $y_0$ is  common to all three 5-branes, 
$y_{1}^{(1)}, y_{2}^{(1)}$ are common  to the second and third 5-branes, etc.
$F_i$ depend on three   $x$-coordinates. The  duality  $*$  is always  defined with respect  to   the transverse $x$-subspace  ($R^3_x$ in \fivv,\ffrof).
The special case of $F_1=F_2=F_3$  gives   the solution 
found in \papd. 
If $F_2=F_3=1$ the above  background reduces to the  
single 5-brane solution \two,\ff\
with the harmonic function $H_5=F_1\inv$ being independent of the two of transverse 
coordinates (here  denoted as $y_{1}^{(1)}, y_{2}^{(1)}$). 
The case of $F_3=1$ describes  two 
5-branes  orthogonally intersecting  over a 3-brane
\eqn\fpvv{
d s^2_{11} =  (F_1 F_2 )^{-2/3}
 \big[ F_1 F_2  (-  dt^2  + dydy_3) 
+ \  F_1  dydy^{(1)}_{2}  + 
  F_2   dydy^{(2)}_{2}  
+  dxdx_{3}\big] \ ,    } 
\eqn\rf{ {{\cal F}_4}_{(5\bot 5)}
 =3( *dF\inv_1 \wedge J_1 +  *dF\inv_2 \wedge J_2 ) \ ,   }
which  again reduces to the  corresponding solution of \papd\  when  $F_1=F_2$. 

The $5\bot 5\bot 5$  configuration \fivv\ has also 
the following     generalisation 
obtained by adding a `boost' along  the common string 
\eqn\fivvw{
d s^2_{11} =  (F_1 F_2 F_3)^{-2/3}
 \big[ F_1 F_2 F_3 (dudv   + K du^2)  } $$
+ \  F_2 F_3 dydy^{(1)}_{2}  + 
  F_1 F_3  dydy^{(2)}_{2} +  F_1 F_2 dydy^{(3)}_{2}  
+  dxdx_{3}\big] \ .     $$
Here $u,v=y_0 \mp t$ and $K$  is a generic harmonic function of the three coordinates $x_s$. 
A non-trivial $K = Q/|x|$ 
describes a momentum flow along the string ($y_0$) direction. 
Upon compactification to $D=4$ along isometric $y_n$-directions
this background reduces \TTT\ to extremal 
dyonic black hole with 
regular horizon which has the same metric as the solution
of \CY.    Thus the  `boosted' $5\bot 5\bot 5$ solution
gives an embedding
 of the 1/8 supersymmetric dyonic black hole in $D=11$ 
which is different from the one discussed in Section 4 below
(see \TTT\ for details). 

\subsec{2-brane intersecting 5-brane }
Let us now  consider other possible  supersymmetric  
intersecting  configurations   not discussed in \papd.
The most important   one is a 2-brane orthogonally intersecting 
a 5-brane over a string
(a possibility of such a configuration was pointed out   in 
\refs{\ast,\ttt}). The corresponding background is 
easily constructed  using the harmonic function rule 
\eqn\vv{
d s^2_{11} =  F^{-2/3} T^{-1/3}
 \big[ F T  (-  dt^2  + dy_1^2) 
 +  F  (dy^2_2 + ... + dy^2_5)    +   T dy_6^2 
+  dxdx_{4}\big] \ ,    } 
\eqn\rfe{ {{\cal F}_4}_{(5\bot 2)}
 = - 3dt\wedge dT \wedge dy_1 \wedge dy_6  + 3 *dF\inv  \wedge dy_6  \ ,   }
where $y_1,...,y_5$  belong to 5-brane and $y_1,y_6$ to 2-brane.
This solution  can be generalised  further 
\eqn\vvw{
d s^2_{11} =  F^{-2/3} T^{-1/3}
 \big[ F T  (dudv + K du^2) 
 +  F  (dy^2_2 + ... + dy^2_5)    +   T dy_6^2 
+  dxdx_{4}\big] \ ,    } 
where as in \fivvw\  $u,v=y_1\mp t$ and $K$,  like $T\inv$ and $F\inv$,  is a generic harmonic function of $x_n$. 
In the simplest 1-center case  having a non-trivial $K$ 
corresponds to adding a momentum flow along the string ($y_1$) direction. 

Dimensional reduction of this solution to $D=10$ along  $x_{11}\equiv y_6$
 (the direction  of  2-brane
orthogonal  to 5-brane)  leads 
to the NS-NS type II   background  corresponding 
to  a fundamental string  lying within a solitonic 5-brane. 
Using the relation between the $D=11$ and (string frame) $D=10$ metrics
\eqn\mett{
ds^2_{11} = e^{4\p/3} \big(dx_{11}^2 +  e^{-2\p} ds_{10}^2\big)  \ , }
we  indeed find  the expected $D=10$ background  with the dilaton 
$e^{2\p} = F\inv T$, the metric  given by \fuss\  (with $H_1=T\inv, \ 
H_5= F\inv$)  and  the antisymmetric 2-tensor field strength  determined by  the  3-tensor field 
strength \rfe.

Dimensional reduction along the string $y_1$ direction leads instead to the $D=10$ solution corresponding to a fundamental string (along $y_2$) 
 orthogonally intersecting a 4-brane (cf. \fof).
Here the dilaton is $e^{2\p} = H_1\inv H_4^{-1/2}, \ H_1= T\inv, \ H_4 = F\inv $
and thus the  resulting  $D=10$  metric 
has indeed the  form \jjj\ 
obtained by  applying the harmonic function rule 
to combine the fundamental string \fus\ and R-R 4-brane \fef\  of type IIA theory.  Another  possibility  is to  compactify along one 
of the transverse directions,  e.g.,  $x_4$ (assuming that harmonic functions 
are independent of it or forming a periodic array) 
in which case we find the type IIA solution describing  a  
R-R  2-brane orthogonally intersecting solitonic 5-brane.

Compactification of  all  6 isometric  $y$-coordinates on a 6-torus
leads to the  extremal $D=5$ black hole solution parametrised by 3
independent charges \TT.  Thus  the `boosted' $2\bot 5$ 
solution and  $2\bot 2\bot 2$ solution  discussed above  represent  two 
different $D=11$ `lifts' of the regular extremal 3-charge  $D=5$ black hole.

These black holes have a finite entropy\foot{This  makes possible  
  to reproduce their entropy  
by counting the corresponding BPS states using D-brane description of the
corresponding dual backgrounds with R-R charges \refs{\SV,\CM} or using
 direct conformal field theory considerations \TT.}
which  is not surprising since   \vvw\ 
has a finite entropy directly as a $D=11$ black brane 
background (assuming that 
internal directions of 2- and 5-branes are compactified). 
Setting $T\inv = 1 + \td Q/r^2, \  F\inv= 1 + P/r^2, \ K=  Q/r^2 \ (r^2= x_mx_m) , $
 one finds that $r=0$ is a regular horizon  (all radii  are regular at $r\to 0$) with the  area
$A_9 = 2\pi^2 L^6 \sqrt{Q \td Q P}$ ($L$ is an equal  period of $y$-coordinates).
The corresponding  thermodynamic entropy can then be understood  as a statistical entropy (related  to  existence of degenerate   $5\bot2$ 
BPS configurations  with the same values of the charges) 
   by counting  relevant   BPS states  directly in  $D=11$ 
as suggested in \dvv.

The $2\bot 5$ metric \vv\ may be compared  to 
the metric 
 obtained by lifting to $D=11$ the  $D=8$ 
dyonic membrane solutions \iz\  
\eqn\izq{d s^2_{11} =  T^{-1/3} \td T^{-1/3}
 \big[  T   (-  dt^2  + dy_1^2 + dy_2^2) 
 +  \td T   (dy^2_3 + dy^2_4 + dy^2_5)  +  dxdx_{5}\big] \ ,  } 
where $T\inv =1 + q/|x|^3, $ and $\td T\inv =1 +\td q/|x|^3$, 
\ $\td q= q \cos^2 \xi $
($\xi$ is  a free parameter). Since \izq\  
  reduces to the 2-brane metric  if $\td T=1$ and to the 
5-brane metric if $\td T =T$  (and thus is similar to  
the metric \fouf\ of  a bound state of a NS-NS and R-R strings
in type IIB theory)  this background can  presumably  be interpreted 
as  corresponding to  a  2-brane lying within a  5-brane
 \refs{\iz,\papd}.

\subsec{$2\bot 2\bot 5$ and $5\bot 5 \bot 2$ configurations}
Two other   1/8  supersymmetric 
configurations of three orthogonally intersecting M-branes
are  $2\bot 2\bot 5$ and  $5\bot 5 \bot 2$. The first one 
represents  two 2-branes  each intersecting 5-brane over a string 
with  the  two  strings intersecting over a point (so that 2-branes 
intersect only over a point).  
The second one  corresponds to  a  2-brane intersecting each of the 
two 5-branes over a string with 
the 5-branes intersecting over a 3-brane (with the strings 
orthogonally intersecting 3-brane over a point). 

In the first case we find
\eqn\tuw{
d s^2_{11} =  (T_1 T_2 )^{-1/3} F^{-2/3} 
 \big[ - T_1 T_2  F  dt^2  } $$
+ \  T_1 F dy_1^2   + T_1  dy^2_2   + 
 T_2 F  dy^2_3 +  T_2 dy_4^2  
 +  F   (dy^2_5 +  dy^2_6  + dy^2_7 )  
+  dxdx_{3}\big] \ ,     $$
\eqn\uff{ {{\cal F}_4}_{(2\bot 2\bot 5)}
 = -3dt\wedge d(T_1 dy_1\wedge d y_2+ T_2 dy_3\wedge d y_4) + 3 *dF\inv  \wedge dy_2\wedge dy_4  \ ,  }
where $y_1,y_3, y_5,y_6,y_7$ are  5-brane coordinates  
and $y_1,y_2$ and $y_3,y_4$ are coordinates of 2-branes.\foot{ 
Note that this configuration is unique
since (according to 
the rule that $p$-branes can intersect only over  $(p-2)$-branes) 
the 2-branes cannot intersect over a string.  
 For  example, if one would try to modify \tuw\ by 
combining  $dy_1^2$ with $dt^2$  then $y_1$ 
would  belong also to the second 2-brane.}
In the second case
\eqn\fivvu{
d s^2_{11} =  T^{-1/3} (F_1 F_2)^{-2/3}
 \big[ - F_1 F_2 T  dt^2  } $$
+ \  F_1 T  dy_1^2 +  F_2 T dy_2^2 +   
 F_1F_2 ( dy_3^2 + dy_4^2 + dy_5^2) 
+ F_1 dy^2_6 + F_2 dy^2_7 
+  dxdx_{3}\big] \ ,     $$
\eqn\ffrofd{ {{\cal F}_4}_{(5\bot 5\bot 2)}
 = -3dt\wedge d(T dy_1\wedge d y_2) +
3( *dF\inv_1 \wedge dy_2 \wedge dy_7 +  *dF\inv_2 
\wedge dy_1 \wedge dy_6 ) \ .  }
Here $y_1,y_2$ belong to  the 2-brane 
and $y_1,y_3,y_4,y_5,y_6$ and $y_2,y_3,y_4,y_5,y_7$ are coordinates of the two 5-branes intersecting over  $y_3,y_4,y_5$.\foot{Another possibility could be  to   consider  2-brane intersecting each of 
the two 5-branes over the same  string,  i.e.  $
d s^2_{11} =  T^{-1/3} (F_1 F_2)^{-2/3}
 \big[ F_1 F_2 T  (-  dt^2  + dy_1^2)  $  $
+ \  T  dy_2^2  +  F_1F_2 (dy_3^2 + dy_4^2)  + 
  F_1 (dy^2_5 + dy^2_6)    +   F_2 (dy^2_7 + dy^2_8) 
+  dxdx_{2}\big]$.
In this case, however, 
the  transverse space  is only 2-dimensional and  thus 
the harmonic functions do not decay  at infinity. }

The backgrounds \tuw,\uff\ and \fivvu,\ffrofd\ 
 have  `dual' structure.
In the special case when $T_2=1$ in \tuw,\uff\
and $F_2=1$ in \fivvu,\ffrofd\ 
they become equivalent to the  $2\bot5$ 
solution \vv,\rfe\ with the harmonic functions
 independent of one of the 
4 transverse coordinates ($y_4$ in \tuw\ and $y_7$ in \fivvu). 
Various possible dimensional reductions to $D=10$ lead to expected 
p-brane intersection configurations of type IIA theory.
For example, the reduction of $2\bot 2\bot 5$ \tuw\ 
along the  orthogonal direction $y_2$ of the first 2-brane  
leads to the configuration of a solitonic 5-brane 
with  a  fundamental string lying within it 
orthogonally intersected by
 2-brane.  Dimensional reduction along the direction $y_1$ 
common to the first 2-brane and 5-brane  leads 
to the  4-brane orthogonally intersected by fundamental string and 2-brane, 
while the reduction along other  5-brane directions ($y_{5},y_6,y_7)$
gives  $2\bot 2\bot 4$  type IIA configuration, etc.

\newsec{$D=11$ solution corresponding to $D=4$ extremal dyonic 
black  hole}
The extreme dyonic  $D=4$ black hole    string solutions 
with non-zero  entropy  \refs{\kall, \CY}  are 
 described by the following  NS-NS  type II 
$D=10$ background (compactified on 6-torus)   \US 
\eqn\mei{
   ds^2_{10}  =  H_1\inv (x) [d u d v + K(x) du^2]
 +  dydy_4   }
$$  +  \   H_5 (x) V\inv (x) [dy_2 + a_s (x) d x^s]^2  + 
H_5 (x)  V (x) dx dx_3    \ , $$ 
\eqn\mii{ e^{2\p} = H_1\inv H_5  , 
\ \  B = H_1\inv  dt \wedge dy_1 - b_s dx^s\wedge dy_2   , 
\ \  db = -*dH_5 ,  \  da = - *dV , } 
where $u,v= y_1 \mp t$ and 
$H_1,H_5, K, V $   are harmonic  functions of $x_s$ ($s=1,2,3$).
This background can be interpreted
 as representing  a  fundamental string (with an extra momentum 
along it, cf. \fuss)  lying within  a 
 solitonic  5-brane  
 with  all 
harmonic functions being independent of one of the four transverse directions ($y_2$)  along which a  Kaluza-Klein 
monopole  \grop\ is introduced.
 Since the corresponding \sm is 
invariant under $T$-duality, one cannot get rid of  the off-diagonal 
KK monopole term in the metric by  dualizing in $y_2$ direction. 
However, 
interpreting this background as a solution of type IIB 
theory one can apply the $SL(2,Z)$ duality 
to transform it first into a configuration  of a  R-R string lying on  a 
R-R 5-brane  
`distorted' by the  Kaluza-Klein monopole. The  metric one finds is then  given by  \mei\ 
rescaled by $e^{-\p}$, i.e. 
\eqn\meiy{
   ds^2_{10 IIB}  = (H_1 H_5)^{1/2} \big[ H_1\inv  H_5\inv  (d u d v + K du^2)
 +  H_5^{-1} dydy_4  } 
$$ 
 +  \      V\inv   (dy_2 + a_s  d x^s)^2 
+    V  dx dx_3 \big]   \ . $$
 Since  $B_{\m\n}$ in \mii\ is transformed into  a R-R field
one  can  now use    $T$-duality along $y_2$ 
to exchange   the  off-diagonal term in the metric 
for an extra NS-NS $B_{\m\n}$ field. The resulting type IIA background has the metric  
 \eqn\mey{
   ds^2_{10 IIA}    = (H_1 H_5)^{1/2} V \big[ H_1\inv  
H_5\inv V\inv (d u d v + K du^2)
   +   H_5^{-1} V\inv dydy_4  }
$$  +  \  H_5^{-1} H_1\inv d\td y_2^2 +
 dx dx_3  \big]  \  ,    $$
and the dilaton $e^{2\p'} = H_1^{1/2} H_5^{-3/2} V$. 
As in  other examples discussed above we can  interpret
this metric as describing  a 
solitonic 5-brane (with the corresponding harmonic function  now being $V$)
which is lying  within a  
R-R 6-brane (with the  harmonic function $H_5$ and an extra 
 dimension  $\td y_2$), both being  orthogonally intersected 
(over a string along $y_1$) 
by  a R-R 2-brane (with coordinates $y_1,\td y_2$  and 
 harmonic function $H_1$).
 Equivalent interpretation of this  $D=4$ dyonic black hole background 
was suggested in \mast\ where it 
 was used to argue  that statistical entropy 
found by  $D$-brane counting 
of  degenerate BPS states reproduces the finite thermodynamic 
entropy  of the black hole.

Anticipating a possibility 
to compute the entropy  by counting  BPS states directly in $D=11$ 
theory 
\dvv\  
it is of interest to lift  the above  type IIA $D=10$ background 
to $D=11$.  Both  forms 
of the $D= 10$ type IIA  solution  \mei\ and \mey\  
lead to equivalent non-diagonal $D=11$ metric.\foot{ 
Though the $D=10$ metric \mey\ is diagonal,  the  R-R vector 
field  supporting the 6-brane  gives a non-vanishing
   $G_{11\m}$ component of the $D=11$ metric.} 
From  \mei,\mii\  we find 
\eqn\meiy{
   ds^2_{11}  = H_1^{1/3} H_5^{2/3} V  \big[  H_1\inv H_5\inv  V\inv 
 (d u d v + K du^2) + 
H_5\inv  V\inv  dydy_4  } $$  +  \ 
   H_1^{-1} V\inv  dx_{11}^2 
 +    V^{-2}    (dy_2 + a_s d x^s)^2  
   +  dx dx_3 \big]     \ .  $$ 
The corresponding 3-tensor field strength  ${\cal F}_4$ is
\eqn\qqq{
{\cal F}_4 =   3 dB \wedge dx_{11} =
- 3 dt \wedge dH_1\inv  \wedge dy_1 \wedge dx_{11} 
 + 3 *dH_5 \wedge dy_2 \wedge dx_{11} \ . }
Starting with  \mey\ one obtains  
 equivalent metric with $ V\leftrightarrow H_5$,\  $x_{11} \to \td y_2$, 
$y_2 \to x_{11}$.
The metric  \meiy\ can be interpreted as  describing  
{\it intersecting 2-brane and 5-brane}  (cf. \vv\  for $V=1$, $F=H_5\inv, 
\ T= H_1\inv$, \ $x_{11}=y_6, \ y_2=x_4$)
 {\it superposed with a  KK monopole }
along $y_2$ (for $H_1=H_5=1,\ K=1$ the metric becomes that of 
KK monopole times a 6-torus  or type IIA  6-brane  lifted to $D=11$, 
see  second 
 reference in \hult).

The special cases of  the background \mei\ 
when one or more harmonic functions are trivial  are related 
to $a=1/\sqrt3, \sqrt 3, 1$ extremal   $D=4$ black holes. The  `irreducible' 
case when 
all 4 harmonic functions  are non-trivial and equal  ($H_1=H_5=K=V$) 
corresponds (for the  1-center choice of $V$) 
to the  $a=0$,   $D=4$ (Reissner-Nordstr\"om)  black hole.
The associated $D=11$ metric \meiy\  takes the  form 
\eqn\muuy{
   ds^2_{11}  =   V\inv (x)  d u d v +  du^2 +   dydy_4  +  dx_{11}^2 
 +     (dy_2 + a_s dx^s)^2  +  V^2(x)  dx dx_3      \ .  } 
We   conclude (confirming
 the expectation  in \papd)
that there exists an embedding of  $a=0$ 
 RN  black hole  into $D=11$ 
theory which has a  non-trivial KK  monopole type metric.
This seems to represent an obstacle on the way of applying 
the $D=11$ approach  \dvv\
in order to give a statistical derivation of the $D=4$ 
black hole entropy:  one is  to understand the 
effect of the presence of the  KK monopole 
 on counting  of BPS states  of systems of M-branes.\foot{For 
 a discussion  suggesting   possible irrelevance 
of KK monopole for counting of BPS states  in the $D$-brane approach 
see \myers.}

The embedding of  extreme 1/8 supersymmetric 
dyonic black holes  into $D=11$ theory discussed above is not, however, 
the only possible one. There exist two  different 1/8 supersymmetric $D=11$ 
solutions, namely, $5\bot5\bot5$ with a `boost' along the common string (Section 3.1) and 
$2\bot2\bot5\bot5$,  for which the $D=11$ metric does not have 
KK monopole part but  still  reduces  to an equivalent 
 $D=4$ dyonic black hole metric with regular horizon and finite entropy \TTT.
These M-brane configurations are likely to be a  proper starting point 
for a statistical understanding of $D=4$ black hole entropy 
directly from  M-theory point of view.

\newsec{Concluding remarks }
As was discussed above, there are 
 simple rules of constructing supersymmetric
composite M-brane  
  solutions from the basic building blocks --  $D=11$ 2-brane  and 5-brane.
 This may be considered as an indication  that  there may  exist
 a $D=11$  analogue   of   D-brane  description  of R-R solitons 
in type II $D=10$
 string theories   which  applies   
 directly 
to  supersymmetric BPS  configurations  of  $D=11$  supergravity 
 (in agreement with related suggestions
in \refs{\ast,\ttt,\beck,\aar,\dvv,\ald}).

We have also presented some  explicit solutions
corresponding to  intersecting p-brane configurations  of 
$D =10$  type II theories. The resulting gravitational 
backgrounds  complement  
the  picture implied by D-brane approach.
An advantage  of viewing 
type IIA $D=10$ configurations from $D=11$ perspective  
is  that this makes possible to treat
various combinations of  NS-NS and R-R p-branes 
on an equal footing, and in this sense   goes  beyond the D-brane
description.

\newsec{ Acknowledgements}
I am grateful to  M. Cveti\v c, I. Klebanov and  G. Papadopoulos for useful correspondence
on related issues 
and 
acknowledge also the support of PPARC,
ECC grant SC1$^*$-CT92-0789 and NATO grant CRG 940870.

\vfill\eject
\listrefs
\end